\documentclass[11pt]{article}
\usepackage{a4p}
\usepackage{amssymb}
\usepackage{epsfig}
\usepackage{cite}
\usepackage{pennames}
\usepackage{rotating}

\def\d0{|d_0|}
\def\z0{|z_0|}

\newcommand{\TGC}{TGC}
\newcommand{\Wpm}{${\rm W}^\pm$}

\newcommand{\WpWm}{${\rm W}^{+}{\rm W}^{-}$}
\newcommand{\emep}{${\rm e}^{+}{\rm e}^{-}$}
\newcommand{\WWen}{${\rm W}{\rm W}\rightarrow {\rm q}{\rm q}e{\nu}$}
\newcommand{\WWmn}{${\rm W}{\rm W}\rightarrow {\rm q}{\rm q}\mu{\nu}$}
\newcommand{\WWtn}{${\rm W}{\rm W}\rightarrow {\rm q}{\rm q}\tau{\nu}$}
\newcommand{\WWln}{${\rm W}{\rm W}\rightarrow {\rm q}{\rm q}\ell{\nu}$}
\newcommand{\WWqqqq}{${\rm W}{\rm W}\rightarrow {\rm q}{\rm q}{\rm q}{\rm q}$}
\newcommand{\W}{${\rm W}$}
\newcommand{\Wm}{${\rm W}^{-}$}
\newcommand{\Wp}{${\rm W}^{+}$}
\newcommand{\tm}{$\tau_{-}$}
\newcommand{\tp}{$\tau_{+}$}
\newcommand{\Zzero}{${\rm Z}^0$}
\newcommand{\eetoWW}{${\rm e}^+{\rm e}^- \rightarrow {\rm W}^+{\rm W}^-$}
\newcommand{\LepII}{LEP2}
\newcommand{\WWZ}{${\rm WWZ}$}
\newcommand{\WWG}{${\rm WW}\gamma$}
\newcommand{\GeV}{GeV}
\newcommand{\ipb}{pb$^{-1}$}
\newcommand{\qqen}{${\rm q}{\rm q}{\rm e}{\nu}$} 
\newcommand{\qqmn}{${\rm q}{\rm q}\mu{\nu}$} 
\newcommand{\qqtn}{${\rm q}{\rm q}\tau{\nu}$} 
\newcommand{\qqln}{${\rm q}{\rm q}\ell{\nu}$} 
\newcommand{\cthw}{$\cos\theta_{\rm W}$}
\newcommand{\cthstar}{$\cos\theta^{*}$}

\newcommand{\etal}{\mbox{\it et al.,}~}
\newcommand {\beq} {\begin{equation}}
\newcommand {\eeq} {\end{equation}}
\newcommand {\bea} {\begin{eqnarray}}
\newcommand {\eea} {\end{eqnarray}}

\newcommand{\PLB}[3]  {Phys.\ Lett.\ \textbf{B#1} (#2) #3}

\newcommand{\EPC}[3]  {Eur.\ Phys.\ J.\ \textbf{C#1} (#2) #3}
\newcommand{\NIMA}[3] {Nucl.\ Instr.\ Meth.\ \textbf{A#1} (#2) #3}

\begin{document}
\begin{titlepage}
\begin{center}{\large   EUROPEAN ORGANIZATION FOR NUCLEAR RESEARCH
}\end{center}\bigskip
\begin{flushright}
       CERN-EP-2000-113   \\ 18  August, 2000
\end{flushright}
\bigskip\bigskip\bigskip\bigskip\bigskip
\begin{center}{\huge\bf Measurement of W boson polarisations\\
 and CP-violating triple gauge  \\
couplings from W$^{+}$W$^{-}$ production at LEP
}\end{center}\bigskip\bigskip
\begin{center}{\LARGE The OPAL Collaboration
}
\end{center}\bigskip\bigskip
\bigskip\begin{center}{\large  Abstract}\end{center}
Measurements are presented of the polarisation of \WpWm\ boson pairs produced 
in \emep\ collisions,
and of CP-violating \WWZ\ and \WWG\ trilinear 
gauge couplings.
The data were recorded by the OPAL experiment at LEP during 1998,
where a total integrated luminosity of 183 \ipb\ was obtained at a 
centre-of-mass energy of 189 GeV.
The measurements are performed through a spin density matrix analysis of 
the \W\ boson decay products. 
The fraction of \W\ bosons produced with  
longitudinal  polarisation was found to be 
$\sigma_{\rm L}/\sigma_{\rm total} = (21.0 \pm 3.3 \pm 1.6)\%$ 
where the first error is statistical and the second systematic.
The joint \W\ boson pair production fractions were found to be 
$\sigma_{\rm TT}/\sigma_{\rm total} = (78.1 \pm 9.0 \pm 3.2) \%$, 
$\sigma_{\rm LL}/\sigma_{\rm total} = (20.1 \pm 7.2 \pm 1.8) \%$  and 
$\sigma_{\rm TL}/\sigma_{\rm total} = (1.8 \pm 14.7 \pm 3.8) \%$.
In the CP-violating trilinear gauge coupling sector we find
$\tilde{\kappa}_{\rm z} = -0.20^{+0.10}_{-0.07}$, 
$g^{\rm z}_{4} = -0.02^{+0.32}_{-0.33}$ and 
$\tilde{\lambda}_{\rm z} = -0.18^{+0.24}_{-0.16}$, where errors include both 
statistical and systematic uncertainties.
In each case the coupling is determined with 
all other couplings set to their Standard Model values except those related
to the measured coupling via $SU(2)_{\rm L}\times U(1)_{\rm Y}$ symmetry. 
These results are consistent with Standard Model expectations.
\bigskip\bigskip\bigskip\bigskip
\bigskip\bigskip
\begin{center}{\large
(To be submitted to Eur. Phys. J. C
)
}\end{center}
\end{titlepage}
\begin{center}{\Large        The OPAL Collaboration
}\end{center}\bigskip
\begin{center}{
G.\thinspace Abbiendi$^{  2}$,
K.\thinspace Ackerstaff$^{  8}$,
C.\thinspace Ainsley$^{  5}$,
P.F.\thinspace {\AA}kesson$^{  3}$,
G.\thinspace Alexander$^{ 22}$,
J.\thinspace Allison$^{ 16}$,
K.J.\thinspace Anderson$^{  9}$,
S.\thinspace Arcelli$^{ 17}$,
S.\thinspace Asai$^{ 23}$,
S.F.\thinspace Ashby$^{  1}$
% ...
}\end{center}\bigskip
\bigskip
$^{  1}$School of Physics and Astronomy, University of Birmingham,
Birmingham B15 2TT, UK
\newline
$^{  2}$Dipartimento di Fisica dell' Universit\`a di Bologna and INFN,
I-40126 Bologna, Italy
\newline
$^{  3}$Physikalisches Institut, Universit\"at Bonn,
D-53115 Bonn, Germany
\newline
% ...
$^{  5}$Cavendish Laboratory, Cambridge CB3 0HE, UK
\newline
% ...
$^{  8}$CERN, European Organisation for Nuclear Research,
CH-1211 Geneva 23, Switzerland
\newline
$^{  9}$Enrico Fermi Institute and Department of Physics,
University of Chicago, Chicago IL 60637, USA
\newline
% ...
$^{ 16}$Department of Physics, Schuster Laboratory, The University,
Manchester M13 9PL, UK
\newline
$^{ 17}$Department of Physics, University of Maryland,
College Park, MD 20742, USA
\newline
% ...
$^{ 22}$Department of Physics and Astronomy, Tel Aviv University,
Tel Aviv 69978, Israel
\newline
$^{ 23}$International Centre for Elementary Particle Physics and
Department of Physics, University of Tokyo, Tokyo 113-0033, and
Kobe University, Kobe 657-8501, Japan
\newline
% ...

\newpage
  
  
\section{Introduction}

We report  measurements of the properties of  
\W\ pair production in \emep\ collisions using data 
recorded by the OPAL detector at LEP at a centre-of-mass energy of 189
\GeV\ with a total integrated luminosity of 183 \ipb. 
We perform a spin density matrix (SDM) analysis~\cite{operate,gounaris} 
of the production and decay 
properties of the \W\ bosons using the semi-leptonic \WWen, 
\WWmn\ and \WWtn\ final states.
Using suitable summations of SDM 
elements we present measurements
of the inclusive production cross-sections for each of the transverse
and longitudinal polarisation states of the \W.
The SDM elements are also sensitive to triple gauge couplings. We
present measurements of CP-violating couplings involving the \Wpm, 
\Zzero\ and photon.

The doubly resonant \eetoWW\ production process proceeds via s-channel 
 \Zzero\ or photon exchange, or via t-channel neutrino exchange, 
collectively known as the CC03 diagrams.
 The s-channel processes contain the \WWZ\ and \WWG\ triple gauge boson 
vertices. The most general Lorentz invariant Lagrangian describing these
vertices~\cite{dieter,bilenky} 
contains 14 independent couplings. 
In order to facilitate measurements the number of parameters is often
reduced to three by assuming $SU(2)_{\rm L}\times U(1)_{\rm Y}$ gauge
invariance and charge conjugation (C) and parity (P) invariance.
The resulting independent couplings are conventionally taken as:
$\Delta\kappa_{\gamma}$ , $\Delta g^{\rm z}_{1}$ and $\lambda$~\cite{lep2}. 
Measurements of these couplings using data recorded at \LepII\ \cite{lep,cosshift,delphi,183,189paper} and
the Tevatron~\cite{tevatron} have been reported elsewhere.

If C, P and CP-invariance are not assumed 
then several additional couplings may be 
present. The CP-violating ones can be taken as 
$\tilde{\kappa}_{V}$ and $\tilde{\lambda}_{V}$ which violate P and conserve
C,
and  $g^{V}_{4}$ which violates C but conserves P ( {\rm V} = \Zzero\ or $\gamma$)~\cite{dieter}. A further parameter, $g^{V}_{5}$,  violates both C and P but conserves 
CP so is not considered here.
$SU(2)_{\rm L}\times U(1)_{\rm Y}$ symmetry imposes the following 
relations~\cite{papa,erato,private} which are assumed in this analysis:
\begin{eqnarray}
\tilde{\kappa}_{\rm z}&=&-\tan^{2}\theta_{\rm w}\tilde{\kappa}_{\gamma}\nonumber\\
\tilde{\lambda}_{\rm z}&=&\tilde{\lambda}_{\gamma}\\
g^{\rm z}_{4} &=& g^{\gamma}_{4}\nonumber
\end{eqnarray}
All of the triple gauge boson couplings (\TGC s) can in principle be  measured
 through 
observations of the \W\ production angular distribution, and  
distributions of the \W\ decay products. 
One way of realising such an analysis is through the spin density matrix
(SDM) approach. In this approach the individual contributions to the \W\ 
production angular distribution arising from each of the different 
possible helicity states of the \W\ bosons can in principle be determined
exclusively.
These exclusive SDM distributions exhibit different behaviour with respect 
to each of the TGCs. The TGCs can therefore be determined from the SDM 
elements in a second step. In W boson pair 
production, the SDM analysis is particularly suited to the extraction of the 
CP-violating 
couplings. Indeed, the imaginary 
parts of the off-diagonal elements of the SDM are completely insensitive to 
the CP-conserving couplings, and will only deviate 
from their Standard Model predictions in the presence of CP-violation at the 
triple gauge boson vertex. 

Before proceeding to the \TGC\ measurements we investigate 
the exclusive production cross-sections for each of the transverse and 
longitudinal polarisations 
of \W\ bosons.
These can be made in the context of the SDM analysis by suitable summations 
of SDM elements which are described in detail in the following section. 
The study of the longitudinal cross-section is particularly interesting
as this degree of freedom of the \W\ only  arises in the Standard Model 
through the electro-weak symmetry breaking mechanism. Previous measurements 
of the proportion of longitudinally polarised W bosons 
produced at \LepII\ have been been reported by the OPAL~\cite{183} and 
L3~\cite{l3} experiments.

In addition we present limits on the CP-violating parameters. 
Previous limits on the CP-violating TGCs  have been obtained at 
the DELPHI experiment using data at 161 GeV and 172 GeV~\cite{delphi}. 
Limits on the CP-violating WW$\gamma$ 
couplings, $\tilde{\kappa}_{\gamma}$ and $\tilde{\lambda}_{\gamma}$, 
have been obtained at the D0 experiment~\cite{D0} from the 
process ${ \rm p}\bar{\rm p} \rightarrow \ell\nu\gamma + {\rm X}$.
Measurements of the neutron electric dipole moment show that the 
electromagnetic interaction is CP-conserving to very high 
accuracy~\cite{neut}.


\section{The Spin Density Matrix}\label{sec:sdm}

The two-particle joint spin density matrix (SDM)~\cite{bilenky} describes 
the contribution of each of the helicity states of 
the \W\ bosons to the \WpWm\ production cross-section.

The amplitude for the production of a \Wm\ with helicity \tm\ (= $-$1,0 or 1)
and a \Wp\ with helicity \tp\ is denoted as
$F^{(\lambda)}_{\tau_{-}\tau_{+}}(s,\cos\theta_{\rm W})$~\cite{operate,bilenky},
where $\lambda$ $(= \pm\frac{1}{2})$ denotes the helicity 
of the e$^{-}$, $s$ is the square of the centre-of-mass energy 
and $\cos\theta_{\rm W}$ is the \W\ production angle in the centre-of-mass frame. 
The SDM ($\rho$) elements are normalised products of these
amplitudes given by~\cite{bilenky}:
\begin{equation}
\rho_{\tau_{-}{\tau^{\prime}}\!\!_{-}\tau_{+}{\tau^{\prime}}\!\!_{+}}(s,\cos\theta_{\rm W}) = 
\frac{\sum_{\lambda}F^{(\lambda)}_{\tau_{-}\tau_{+}}(F^{(\lambda)}_{\tau_{-}^{
\prime}\tau_{+}^{\prime}})^{*}}{\sum_{\lambda\tau_{+}\tau_{-}}|F^{(\lambda)}_{
\tau_{-}\tau_{+}}|^{2}}
\label{eq:exprho}
\end{equation}
The normalisation ensures that 
$\sum_{\tau_{-}\tau_{+}} \rho_{\tau_{-}\tau_{-}\tau_{+}\tau_{+}} = 1$.
The $\rho$ matrix is Hermitian, giving 80 independent real coefficients 
which may be experimentally measured.
The diagonal elements are defined as the subset of elements where 
$\tau_{-}={\tau^{\prime}}\!\!_{-}$ and  $\tau_{+}={\tau^{\prime}}\!\!_{+}$. 
These diagonal elements are purely real and are equivalent to
the probability of producing a final \WpWm\ state with helicities 
$\tau_{+}$ and $\tau_{-}$ respectively. 
The off-diagonal elements represent interference terms between the different 
helicity states of each \W. 

In the context of a limited sample of events it may not be possible to measure
all of the components independently. It is then useful to consider
the single particle density matrix for either the \Wm\ or \Wp, formed by 
summation over the helicity states of the other \W. For example the 
\Wm\ matrix elements 
$\rho^{W^{-}}_{\tau_{-}{\tau^{\prime}}\!\!_{-}}$ are obtained by summation over
 $\tau_{+}={\tau^{\prime}}\!\!_{+}$: 
\begin{equation}
\rho^{W^{-}}_{\tau_{-}{\tau^{\prime}}\!\!_{-}}(s,\cos\theta_{\rm W}) = \sum_{\tau_{+}}\rho_{\tau_{
-}{\tau^{\prime}}\!\!_{-}\tau_{+}\tau_{+}}(s,\cos\theta_{\rm W}) 
\end{equation}
The Hermitian matrix $\rho^{W^{-}}$ has eight independent real
coefficients, and satisfies the normalisation constraint $\sum_{\tau_{-}}
\rho^{W^{-}}_{\tau_{-}\tau_{-}}=1$. 
The three real diagonal elements represent the relative
production probabilities for a final state \Wm\ with a particular helicity.

The constraints of CPT and CP-invariance impose additional symmetries on
the density matrix at tree level~\cite{gounaris}. CPT-invariance imposes the
conditions:
\begin{eqnarray}
{\rm Re}(\rho^{W^{-}}_{\tau_{1}\tau_{2}}) &-& {\rm Re}(\rho^{W^{+}}_{-\tau_{1}-\tau_{2}}) = 0
\label{eq:stuff}\\
{\rm Im}(\rho^{W^{-}}_{\tau_{1}\tau_{2}}) &+& {\rm Im}(\rho^{W^{+}}_{-\tau_{1}-\tau_{2}}) = 0\label{eq:stuff2}
\end{eqnarray}
CP invariance imposes the condition:
\begin{eqnarray}
{\rm Im}(\rho^{W^{-}}_{\tau_{1}\tau_{2}}) &-& {\rm Im}(\rho^{W^{+}}_{-\tau_{1}-\tau_{2}}) = 0
\label{eq:test}
\end{eqnarray}
Thus CPT and CP-conservation together dictate that all 
coefficients from $\rho^W_{\tau_{1}\tau_{2}}$ are real. 
Deviations from the validity of equation (\ref{eq:test}) would represent 
an unambiguous signal for CP-violation at tree level. Deviations from the 
equality of equation (\ref{eq:stuff2}) represent a signal of higher-order (loop)
 effects
beyond tree level~\cite{gounaris}.

The exclusive differential cross-sections for the production of a
\W\ with transverse (T) or longitudinal (L) helicity are obtained by 
weighting the total differential cross-section by the relevant elements of 
the single particle $\rho$ matrix:
\begin{eqnarray}
\frac{{\rm d}\sigma_{\rm T}}{{\rm d}\!\cos\theta_{\rm W}} &=
& (\rho_{++} + \rho_{--}) 
\frac{{\rm d}\sigma}{{\rm d}\!\cos\theta_{\rm W}}\nonumber\\
\frac{{\rm d}\sigma_{\rm L}}{{\rm d}\!\cos\theta_{\rm W}} &=
& (\rho_{00})\frac{{\rm d}\sigma}{{\rm d}\!\cos\theta_{\rm W}}\label{eq:singlew}
\end{eqnarray}
and the corresponding total cross-sections are given by:
\begin{eqnarray}
\sigma_{\rm T} &=& \int^{+1}_{-1}(\rho_{++} + \rho_{--}) \frac{{\rm d}\sigma}{{\rm d}\!\cos\theta_{\rm W}}{\rm d}\!\cos\theta_{\rm W}\nonumber\\
\sigma_{\rm L} &=& \int^{+1}_{-1}(\rho_{00})\frac{{\rm d}\sigma}{{\rm d}\!\cos\theta_{\rm W}}{\rm d}\!\cos\theta_{\rm W}.\label{eq:totx1}
\end{eqnarray}
The single \W\ SDM elements can be obtained from measurements
of the properties of the \W\ decay products by the application of 
suitable projection operators $\Lambda_{\tau\tau^{\prime}}$~\cite{operate,gounaris} which assume the 
V$-$A coupling of the \W\ to fermions. The \W\ decays are characterised by
the polar and azimuthal angles of the decay fermion in the \W\ rest
frame,  $\theta^{*}$ and $\phi^{*}$ respectively.
In this analysis we consider only the semi-leptonic event 
class, \WWln, where one \W\ decays to a lepton (e, $\mu$ or $\tau$) and a 
neutrino  and the other to two hadronic jets. 
In this case 
the values of $\theta^{*}$ and $\phi^{*}$ of the lepton may be determined 
unambiguously and the corresponding single \W\ SDM elements are given by:
\begin{eqnarray}
&&\frac{{\rm d}\sigma({\rm e}^+ {\rm e}^- \rightarrow {\rm W}^+ {\rm W}^-)}{{\rm
d}\!\cos\theta_{\rm W}}
\rho^{W^{-}}_{\tau\tau'} \nonumber\\
&& = \frac{1}{{\rm Br}({\rm W}^{-} \rightarrow \ell^{-} \bar{\nu})}\int
\frac{{\rm d}\sigma({\rm e}^+ {\rm e}^- \rightarrow W^+ \ell^{-}\bar{\nu})}
{{\rm d}\!\cos\theta_{\rm W}{\rm d}\!\cos\theta^{*}{\rm d}\phi^{*}}
\Lambda_{\tau\tau'}(\theta^{*},\phi^{*}){\rm d}\!\cos\theta^{*}{\rm d}
\phi^{*} \label{eq:tfdcs}
\end{eqnarray}
In section \ref{sec:exp} we describe how this expression is realised as a 
sum over the events observed in the data sample.

In the case of the hadronically decaying \W\ 
differentiation between the particle and anti-particle 
decay products is very difficult. However, certain projection operators are 
symmetric under the transformation 
$\cos\theta^{*} \rightarrow -\cos\theta^{*}$, 
$\phi^{*} \rightarrow \phi^{*} + \pi$,~ i.e. the interchange of one of the decay
particles from the W boson with the other decay particle, for example the u-type
quark with the d-type in a hadronically decaying W boson. This means that
a number of combinations of the SDM elements 
may still be extracted from ${\rm W} \rightarrow {\rm q}\bar{\rm q}$ decays: 
these include
$\rho_{++} + \rho_{--}$ and $\rho_{00}$. Both \W\ bosons  in the event can therefore be used to measure the polarised cross-section,
since only these terms appear in equations 
(\ref{eq:singlew}).  

Returning to the full two particle SDM, the analogous expressions to 
(\ref{eq:singlew}) but describing the differential cross-section for 
both W bosons in the pair
being transversely polarised (TT), both being longitudinally polarised (LL), 
or one of each polarisation (TL) are:
\begin{eqnarray}
\frac{{\rm d}\sigma_{\rm TT}}{{\rm d}\!\cos\theta_{\rm W}} &=
& (\rho_{++++} + \rho_{++--} + \rho_{--++} + \rho_{----})\frac{{\rm d}\sigma}{{\rm d}\!\cos\theta_{\rm W}}\nonumber\\
\frac{{\rm d}\sigma_{\rm LL}}{{\rm d}\!\cos\theta_{\rm W}} &=
& (\rho_{0000})\frac{{\rm d}\sigma}{{\rm d}\!\cos\theta_{\rm W}}\label{eq:bothw}\\
\frac{{\rm d}\sigma_{\rm TL}}{{\rm d}\!\cos\theta_{\rm W}} &=
& (\rho_{++00} + \rho_{--00} + \rho_{00++} + 
\rho_{00--})\frac{{\rm d}\sigma}{{\rm d}\!\cos\theta_{\rm W}}\nonumber
\end{eqnarray}
and the analogous expression to (\ref{eq:tfdcs}) is
\begin{eqnarray}\label{eq:5fold}
&&\frac{{\rm d}\sigma({\rm e}^+ {\rm e}^- \rightarrow {\rm W}^+ {\rm W}^-)}{{\rm
d}\!\cos\theta_{\rm W}}
\rho_{\tau_{-}\tau_{-}^{\prime}\tau_{+}\tau_{+}^{\prime}} \nonumber\\
&=&\frac{1}{{\rm Br}({\rm W} \rightarrow {\rm q}{\rm q}){\rm Br}({\rm W} \rightarrow \ell {{\nu}})} \int
\frac{{\rm d}\sigma({\rm e}^+ {\rm e}^- \rightarrow {\rm q}{\rm q}\ell{\nu} )}
{{\rm d}\!\cos\theta_{\rm W}{\rm d}\!\cos\theta^{*}_{-}{\rm d}\phi^{*}_{-}{\rm d}\!\cos\theta^{*}_{+}{\rm d}\phi^{*}_{+}}\\
\nonumber\\
&\times& \Lambda_{\tau_{-}{\tau^{\prime}}\!\!_{-}}(\theta^{*}_{-},\phi^{*}_{-})\Lambda_{\tau_{+}{\tau^{\prime}}\!\!_{+}}(\theta^{*}_{+},\phi^{*}_{+}) {\rm d}\!\cos\theta^{*}_{-}{\rm d}
\phi^{*}_{-}{\rm d}\!\cos\theta^{*}_{+} {\rm d}
\phi^{*}_{+}\nonumber
\end{eqnarray}
where the integral is now over the decay angles of both  W bosons.

\section{The Data Sample and Monte Carlo Simulated Events}

\subsection{The Data Sample}
\noindent
The W-pair data used in this analysis were collected by the OPAL~\cite{Ahmet:1991eg} detector at LEP. 
The accepted integrated luminosity in 1998, evaluated 
using small angle Bhabha scattering events observed in the silicon
tungsten forward calorimeter~\cite{Anderson:1994ve}, is 183.14 $\pm$ 0.55~pb$^{-1}$~\cite{opalxsec}. The luminosity-weighted mean centre-of-mass energy for the data sample
is $\sqrt{s} = 188.64 \pm 0.04$ GeV. 

In this analysis we use only the W-pair events decaying to 
the \qqen ,~\qqmn\ and \qqtn\ channels. 
These \qqln\ events were first selected using the 
likelihood selection described 
in~\cite{183,opalxsec,Ackerstaff:1998fr}.
The selection is designed to optimise the rejection of
${\rm Z}^{0}/\gamma \rightarrow {\rm q}\bar{\rm q}$
and  four-fermion backgrounds. 
A total of 1252 \qqln\ candidates was selected at this stage. Monte Carlo
studies show this selection is about 88\% efficient.

Further cuts are applied
in order to obtain a sample of well reconstructed events. A full description 
can be found 
in~\cite{189paper}. A brief overview of the procedure is given here.
For the \qqen\ and \qqmn\ events a well reconstructed lepton track is required.
 For the 
\qqtn\ events, either one track or a narrow jet consisting of three tracks is assigned as the $\tau$ decay product.

Kinematic fits are now applied to the events. For the \qqen\ and \qqmn\ events
a one-constraint fit is applied that requires energy-momentum conservation
and neglects any initial state radiation. 
Events are accepted if the fit converges with a probability larger than 0.001. 
An  improvement is made to the  resolution of the angular observables by performing a 
second  kinematic fit which constrains each reconstructed W mass to the average
value measured at the Tevatron~\cite{massofw}. Events for
which the fit converges with a probability of at least 0.001 are accepted; for other events
we revert to the previous fit.
 
For the \qqtn\ events a different kinematic fit is applied~\cite{183};
 here the 
reconstructed mass of the two W bosons is required to be equal. In order to 
obtain a W mass
from the leptonic part of these events, it is necessary to assume that the 
direction of the visible part of the $\tau$ decay approximates the direction of 
the $\tau$ lepton. The fit is required to
converge with a probability of at least 0.025. This cut rejects 41\% of the 
background. It also rejects about
      14\% of the signal events, but preferentially rejects 
      events where the $\tau$ decay products are not correctly identified.

\begin{figure}[H]
\begin{center}
\epsfig{file=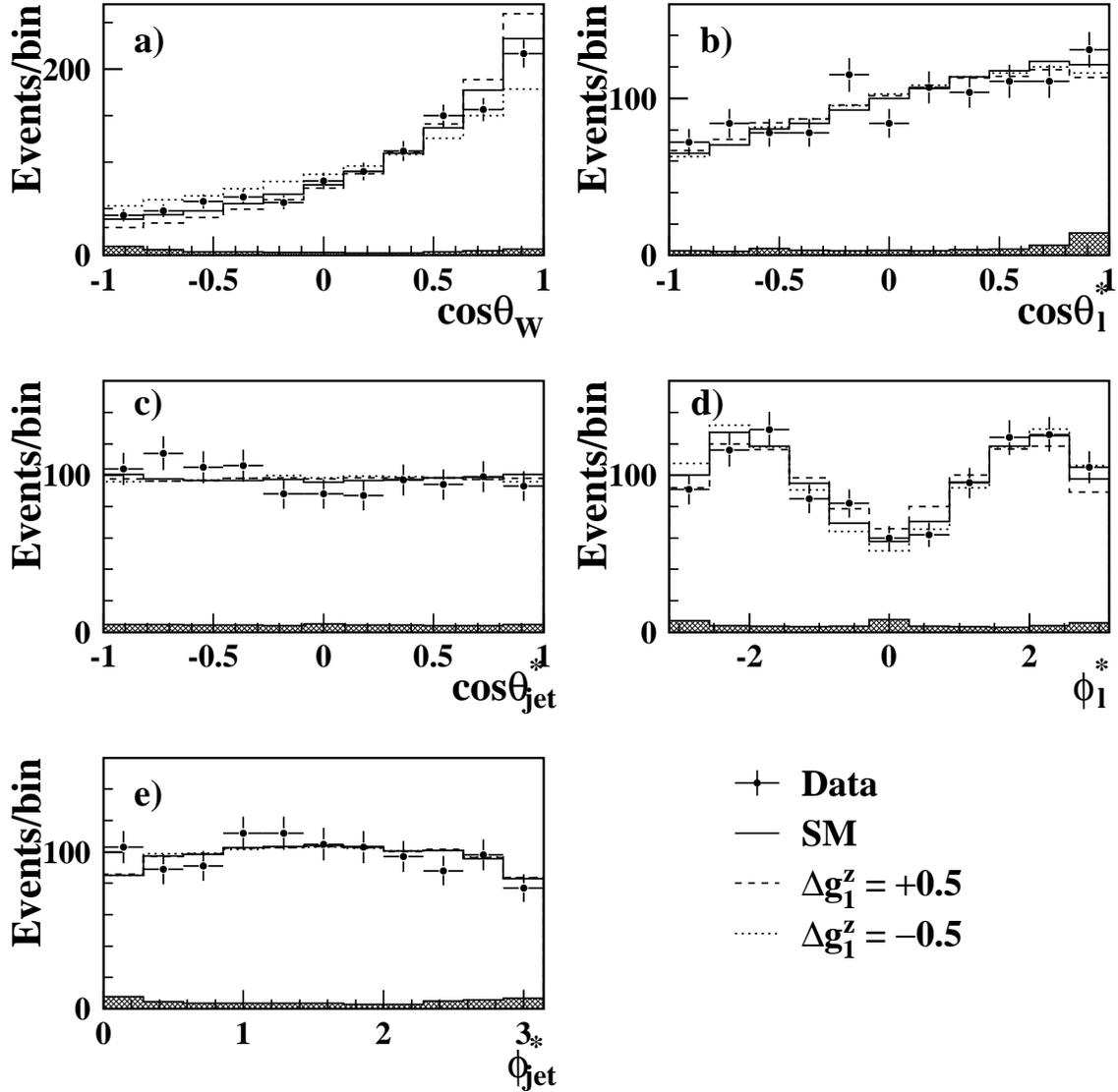,width=1.0\linewidth}
\end{center}
\caption[stuff]{Distributions of the kinematic variables $\cos\theta_{\rm W}$,
$\cos\theta^{*}_{l}$, $\cos\theta^{*}_{\rm jet}$, $\phi^{*}_{l}$ and 
$\phi^{*}_{\rm jet}$, as obtained from the \qqln\ events. The  points represent the data. The histograms show the expectation of the Standard Model and the cases $\Delta g^{\rm z}_{1} = \pm0.5$. The shaded histogram shows the non-\qqln\ background. In the case of the W$^{+}$ decaying to the lepton, the value of 
$\phi^{*}_{l}$ is shifted by $\pi$ in order to overlay the W$^{+}$ and W$^{-}$ distributions. }
\label{fig:angles}
\end{figure}

After application of all selection cuts 1065 \qqln\
candidates remain, with 359 in the \qqen\  channel, 386 in the \qqmn\ channel and 320 in the \qqtn\ channel. The W production and decay angles for these 
events are shown in figure~\ref{fig:angles}.
The background contributions to the data sample are estimated using Monte Carlo
samples. The resulting contaminations are expected to be four-fermion (non-CC03)
3.1\%, ${\rm Z}^{0}/\gamma \rightarrow {\rm q}\bar{\rm q}$ 2.1\%, two-photon (multiperipheral) 0.2\% and CC03 four-fermion
(\WWqqqq\ and $\ell{\nu}{\ell}\nu$) 0.5\%.

\subsection{Monte Carlo}
\noindent
\label{sec:mc}
A number of Monte Carlo models are
used to provide estimates of efficiencies and backgrounds as well as
the expected W-pair production and decay angular distributions for different 
TGC values. The Monte Carlo samples are generated at 
$\sqrt{s} = 189$~\rm{GeV}.
All Monte Carlo samples mentioned below are passed through the full 
OPAL simulation program~\cite{GOPAL} and then subjected to the same 
selection and reconstruction procedure as the data. 
The ERATO~\cite{erato} Monte Carlo program is used in this analysis because it can generate 
samples of four-fermion Monte Carlo events with non-Standard Model values of 
CP-violating anomalous couplings. The EXCALIBUR~\cite{excalibur} Monte Carlo is
 also used
extensively in this analysis  for correction of detector effects and 
as the main tool to compare to the 189 GeV data. A comparison between 
ERATO and EXCALIBUR angular distributions and spin density matrix elements was
 undertaken and  no statistically significant differences were seen between 
them. The other four-fermion generator used to calculate background 
contributions
and systematic uncertainties is grc4f~\cite{grc4f}. The WW generators 
 used to calculate theoretical predictions for the W-pair polarisations and to
calculate systematic uncertainties are EXCALIBUR, HERWIG~\cite{herwig}, 
PYTHIA~\cite{pythia} and 
KORALW~\cite{koralw}. The background ${\rm Z}^{0}/\gamma \rightarrow {\rm q}\bar{\rm q}$ samples are generated using PYTHIA.   For samples of two-photon 
(multiperipheral) background PYTHIA, 
PHOJET~\cite{phojet} and HERWIG were used.

\section{Measurement of W Boson Polarisation}
\label{sec:exp}
\subsection{Experimental Method}
Measurements of 
the production cross-sections for each of the transverse and 
longitudinal states of the \W\ bosons are obtained from summations of
the SDM elements obtained from the \qqln\ data sample, 
after correcting for detector acceptance, resolution effects and
background contamination.
 
To calculate the SDM elements the events are divided into eight equal bins, 
$k$, of \cthw. 
In each bin the SDM elements are obtained by a summation over 
events, $i$, weighted by the relevant SDM projection operators as shown in 
equation~(\ref{eq:proj}). This is 
effectively the realisation of
equation~(\ref{eq:tfdcs}) as a summation over observed events. $N_{k}$ is the 
number of events in bin $k$. In the extraction of all the individual W SDM elements CPT invariance is
assumed, so all leptonic (hadronic) W$^+$ and W$^-$ decays can be combined.
\begin{equation}
\rho^{k}_{\tau\tau'}( \cos\theta^{k}_{W} ) =
\frac{1}{N_{k}}\sum_{i=1}^{N_{k}} 
\Lambda_{\tau\tau'}(\cos\theta_{i}^{*},\phi_{i}^{*})\label{eq:proj}
\end{equation}
Estimates of the production cross-sections are derived from 
the diagonal SDM elements; therefore
the projection operators only involve the polar decay angle \cthstar. 
It is not required that the SDM
elements be positive definite.
Both the leptonic and 
hadronic \W\ decays can be used since the SDM combinations occurring in 
equation
(\ref{eq:singlew}) are symmetric with respect to the use of the 
fermion or anti-fermion polar decay angle.
Each event is also 
weighted by a correction factor, $f$, for detector acceptance 
and resolution effects:
\begin{equation}
\rho^{k}_{00} =
\frac{1}{N^{\rm cor}_{k}}\sum_{i=1}^{N_{k}}
\frac{1}{f_{k}(\theta_{i}^{*})}
\Lambda_{00}(\theta_{i}^{*}) 
\end{equation}
\begin{equation}
\rho^{k}_{++} + \rho^{k}_{--} =
\frac{1}{N^{\rm cor}_{k}}\sum_{i=1}^{N_{k}}
\frac{1}{f_{k}(\theta_{i}^{*})}
( \Lambda_{++}(\theta_{i}^{*}) +
  \Lambda_{--}(\theta_{i}^{*}) 
)
\end{equation}
where $N^{\rm cor}_{k}$ is the corrected number of events in bin $k$:
\begin{equation}
N^{\rm cor}_{k} = \sum_{i=1}^{N_{k}}
\frac{1}{f_{k}(\theta_{i}^{*})}.\label{eq:ncor}
\end{equation}
The factors $f_{k}(\theta_{i}^{*})$ are obtained from fully
    simulated Monte Carlo events, and are calculated as a function of
    \cthw\ and \cthstar. They are defined to be the ratio of
    the number of reconstructed events to the number of generated events
    in each bin. The correction factors have an effect of between 5\% 
    and 20\%. The typical resolution of the measured $\cos\theta_{\rm W}$ is 
found to be 0.05, which is less than 20\% of the bin width used in correcting 
 this distribution. The typical resolution of the measured 
$\cos\theta^{*}$ is found to be 0.07, which is 70\% of the bin width used
in the calculation of the correction factors.

The distribution of uncorrected events, shown in figure~\ref{fig:angles}, 
 also has to be corrected for detector acceptance and resolution effects. 
This is done in a similar 
manner to the correction of the SDM elements by calculating a correction factor
from  Monte Carlo events in a bin-wise fashion and applying it to the data.
The estimated background contribution is subtracted from the SDM elements
and the ${\rm d}\sigma/{\rm d}$\cthw\ distribution.

The polarised differential cross-sections,
${\rm d}\sigma_{\rm L}/{\rm d}\!\cos\theta_{\rm W}$
and
${\rm d}\sigma_{\rm T}/{\rm d}\!\cos\theta_{\rm W}$,  are then obtained
by multiplying the binned unpolarised differential cross-section by the 
relevant SDM combinations, following equations~(\ref{eq:singlew}).

\subsection{Experimental Results}

\begin{figure}[H]
\begin{center}
\epsfig{file=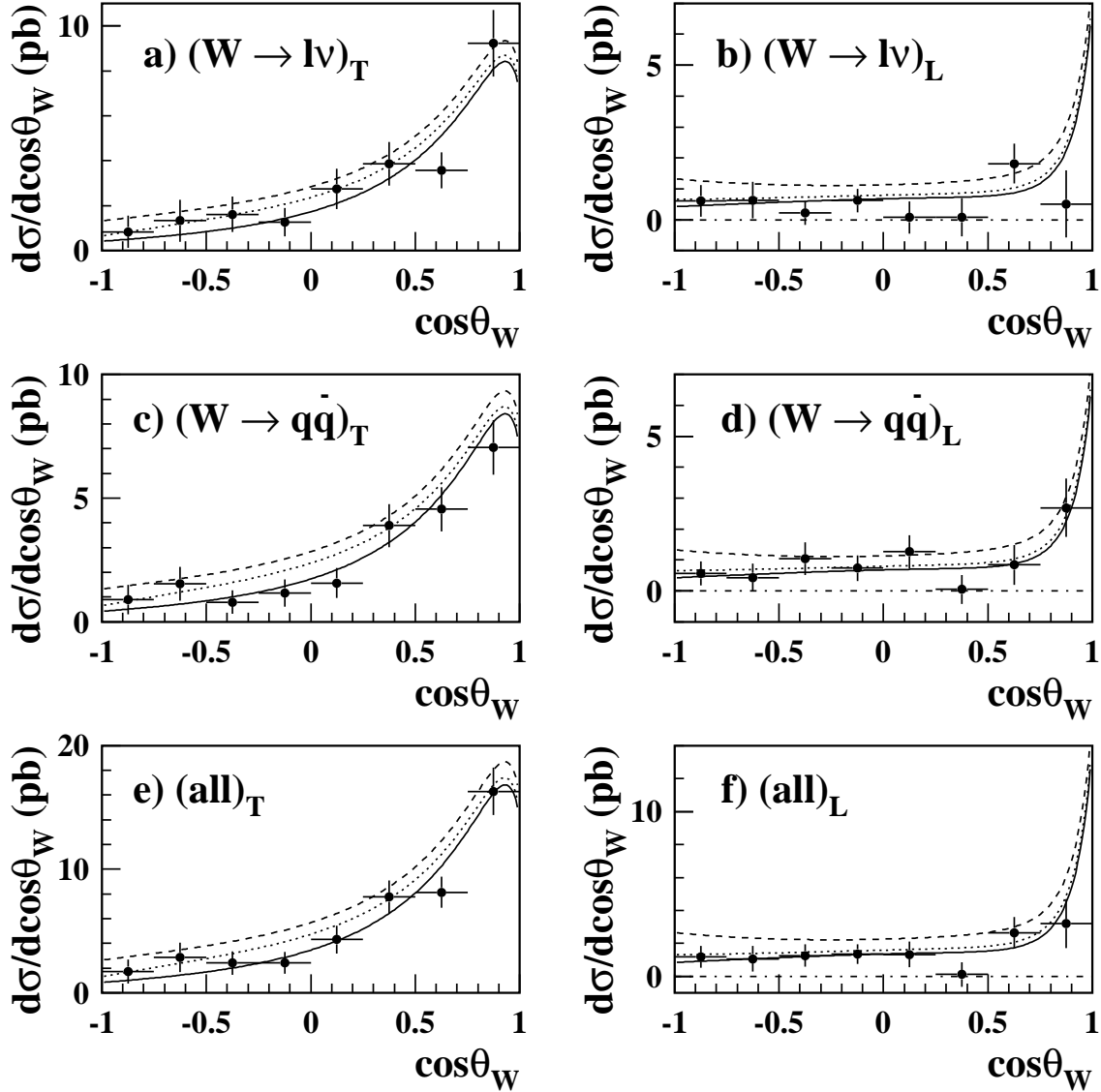,width=1.0\linewidth}
\end{center}
\caption[stuff]{The polarised differential W production cross-sections. a) is 
the differential cross-section for transversely polarised leptonically decaying 
W bosons in the W boson pair and b) is for longitudinally polarised. c) and 
d) are as a) and b) but are shown for the hadronically decaying W boson in the pair. 
e) and f) are the combinations of the leptonically and 
hadronically decaying W. Overlaid are the predictions for the Standard Model 
(solid line) and CP-violating anomalous 
couplings $\tilde{\lambda}_{\rm z} = -0.5$ (dotted line), 
$\tilde{\kappa}_{\rm z} = +0.5$ (dashed line). The dotted-dashed line in b) and f) shows the zero. The errors include both statistical and systematic 
uncertainties. }
\label{fig:pollep2}
\end{figure}

Figure~\ref{fig:pollep2} shows the individual W boson differential 
cross-sections obtained from the data by this method. 
The transverse and longitudinal components obtained from the leptonically 
decaying W and the hadronically decaying W are shown separately, together
with the values obtained by combining the two. The polarisations of the W bosons
 in the W-pair event are correlated. The correlation is estimated to be 0.07,
and is taken into account in the errors on the combined 
cross-sections.

The fraction of each polarisation state, obtained  by integrating over $\cos\theta_{\rm W}$ and then dividing by the total cross-section, is
given in table~\ref{table:hel}. Once again the correlated polarisations of the two W bosons are taken into account when deriving the errors. The estimation of
the systematic errors shown in tables~\ref{table:hel} and~\ref{table:hel2} is described in section~\ref{sec:systematics}.
 \begin{center}
\begin{table}[ht]
\begin{center}
\begin{tabular}{|l|c|c|} \hline
 & $\sigma_{\rm T}/\sigma_{\rm total}$ &  $\sigma_{\rm L}/\sigma_{\rm total}$  \\\hline
\underline{Data} & &   \\
W$ \rightarrow \ell{\nu}$ & 0.842 $\pm$ 0.048 $\pm$ 0.023 & 0.158 $\pm$ 0.048 $\pm$ 0.023\\
W$\rightarrow {\rm q}{\rm q}$ & 0.738 $\pm$ 0.045 $\pm$ 0.025 & 0.262 $\pm$ 0.045 $\pm$ 0.025\\
All  & 0.790 $\pm$ 0.033 $\pm$ 0.016 & 0.210 $\pm$ 0.033 $\pm$ 0.016\\\hline
\underline{Standard Model Expectation} & &  \\
W$\rightarrow \ell{\nu}$ &  0.746 $\pm$ 0.006 & 0.254 $\pm$ 0.006 \\
W$\rightarrow {\rm q}{\rm q}$ & 0.741 $\pm$ 0.006 & 0.259 $\pm$ 0.006\\
All  & 0.743 $\pm$ 0.004 & 0.257 $\pm$ 0.004\\\hline
\end{tabular}
\caption{The fractions of transversely and longitudinally polarised W bosons. The expected values are from generator level EXCALIBUR Monte Carlo, the errors being statistical only. The first error on the measured values is statistical and the second is the systematic uncertainty.}
\label{table:hel}
\end{center}
\end{table}
\end{center}

\begin{figure}[H]
\begin{center}
\epsfig{file=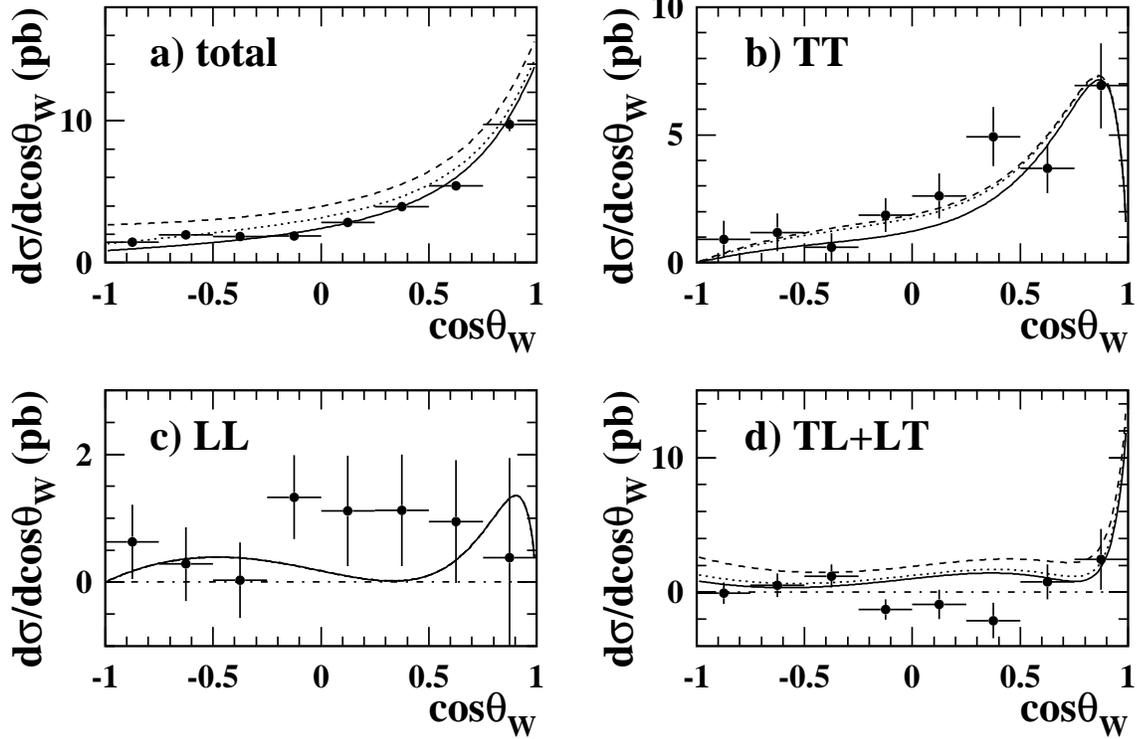,width=1.0\linewidth}
\end{center}
\caption[stuff]{The differential cross-sections for the W bosons of different helicity states.  a) is the corrected total W production differential cross-section, b) is for production of pairs of transversely polarised W bosons, c) is pairs of longitudinally polarised W bosons and d) is for one of each. Overlaid are the predictions for the Standard Model (solid line) and CP-violating anomalous 
couplings $\tilde{\lambda}_{\rm z} = -0.5$ (dotted line), $\tilde{\kappa}_{\rm z} = +0.5$ (dashed line). Only the solid line is visible on plot c) because this distribution is insensitive to the CP-violating couplings. The dotted-dashed line on plots c) and d) shows the zero. The errors include both statistical and systematic uncertainties. }
\label{fig:pollep1}
\end{figure}

The joint polarised cross-sections are obtained
by extracting the joint SDM elements in a similar way to equations~(\ref{eq:proj}) 
\begin{equation}
\label{eq:proj2}
\rho_{\tau_{-}\tau_{-}^{\prime}\tau_{+}\tau_{+}^{\prime}}^{k} =
\frac{1}{N^{cor}_{k}}\sum_{i=1}^{N_{k}}
\frac{1}{f_{k}({\theta_{i}^{*}}^{\rm lept},{\theta_{i}^{*}}^{\rm had})}
\Lambda_{\tau_{-}{\tau^{\prime}}\!\!_{-}}^{\rm W^{\pm}}({\theta_{i}^{*}}^{\rm lept}) 
\Lambda_{\tau_{+}{\tau^{\prime}}\!\!_{+}}^{\rm W^{\mp}}({\theta_{i}^{*}}^{\rm had}) 
\end{equation}
where in this case the $\rho$ and $\Lambda$ are those corresponding to the
particular SDM combinations appearing in equation~(\ref{eq:bothw}).
Note that there is now a projection operator for each \W, and the 
correction factor is binned in terms of both polar decay angles.
Figure~\ref{fig:pollep1} shows the joint differential cross-sections 
obtained by this method, these cross-sections are not constrained to be 
positive definite.

The fractions of each helicity state, obtained by integrating over 
$\cos\theta_{\rm W}$ of the W and then dividing by the total cross-section, 
are shown in table~\ref{table:hel2}. These results
are highly correlated. The correlations between each helicity fraction, calculated from the data,  are 
found to be: $\sigma_{\rm TT}$:$\sigma_{\rm LL}$ = 0.67$\pm$0.02, $\sigma_{\rm TT}$:$\sigma_{\rm TL}$ = $-$0.93$\pm$0.01 and $\sigma_{\rm LL}$:$\sigma_{\rm TL}$ = $-$0.89$\pm$0.01.

The individual W  polarised differential cross-sections show good agreement with 
the Standard Model predictions, as do the overall fraction of each 
individual W polarisation, as seen in table~\ref{table:hel}. 
The W pair polarised cross-sections show less good agreement, but  $\sigma_{\rm LL}/\sigma_{\rm total}$ is
within two standard deviations of the Standard Model prediction and the other two are just over two standard deviations away. The $\chi^{2}$
value for the three measurements compared to the Standard Model expectations,
including systematic uncertainties, is 4.7. This corresponds to a $\chi^{2}$
probability of 10\%.

\begin{center}
\begin{table}[ht]
\begin{center}
\begin{tabular}{|l|l|l|} \hline
 &  Measured  & Expected  \\\hline
$\sigma_{\rm TT}/\sigma_{\rm total}$ & 0.781 $\pm$ 0.090 $\pm$ 0.033 & 0.572 $\pm$ 0.010 \\\hline
$\sigma_{\rm LL}/\sigma_{\rm total}$ & 0.201 $\pm$ 0.072 $\pm$ 0.018  & 0.086 $\pm$ 0.008 \\\hline
$\sigma_{\rm TL}/\sigma_{\rm total}$ & 0.018 $\pm$ 0.147 $\pm$ 0.038 & 0.342 $\pm$ 0.016 \\\hline
\end{tabular}
\caption{The fraction of each helicity combination of WW pairs.  The expected values are calculated from Monte Carlo studies. The first errors on the measured fractions are statistical and the second are systematic. The errors on the expected fractions are statistical only.}
\label{table:hel2}
\end{center}
\end{table}
\end{center}

\section{Measurement of Anomalous Couplings and Test of CP-Invariance}

Measurements of anomalous coupling parameters are obtained 
through a comparison of distributions of SDM elements 
obtained from the leptonically decaying W bosons  in the 
\qqln\ data sample 
with distributions obtained from 
fully simulated Monte Carlo samples. 
In contrast to the polarised cross-section measurements, 
neither data nor Monte Carlo events are corrected for detector 
or acceptance effects. Instead the experimentally observed SDM distributions 
are compared directly with those for fully detector simulated Monte Carlo 
events.
Monte Carlo samples for arbitrary \TGC\ values are obtained by a
re-weighting technique applied to a large Standard Model sample. 
A $\chi^2$ minimisation procedure is then used to
find the simulated data set which best fits the data and hence obtain
the best fit \TGC\ parameter value. 

\begin{figure}[H]
\begin{center}
\epsfig{file=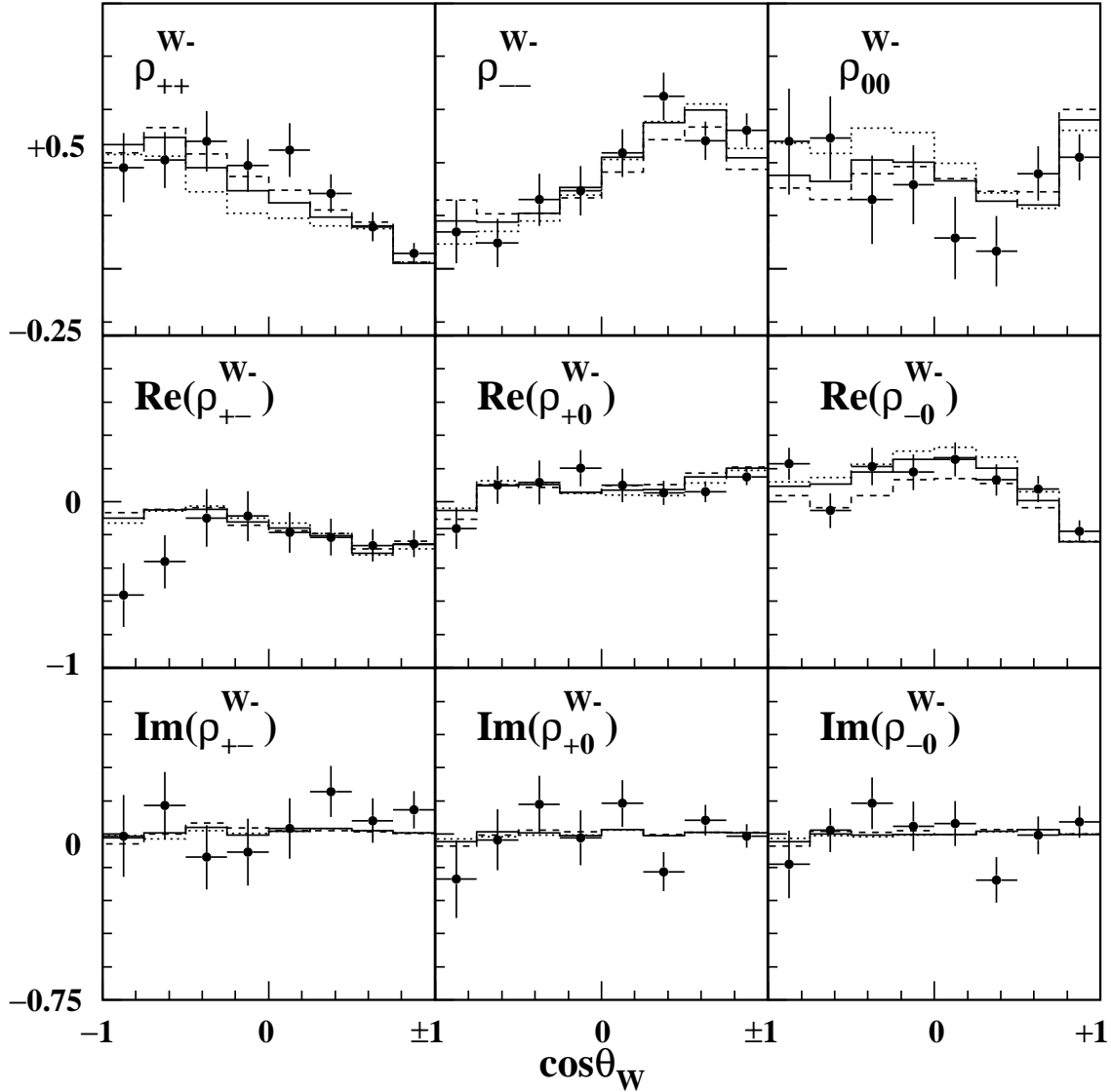,width=1.0\linewidth}
\end{center}
\caption[stuff]{The individual W SDM elements extracted from the leptonically decaying W boson in the \qqln\ data events at 189 GeV. The points are the OPAL data
. The histograms represent the Monte Carlo predictions calculated from fully 
detector simulated Monte Carlo. The solid line shows the Standard Model expectation and the dotted (dashed) line that for $\Delta g^{\rm z}_{1}$ = +0.5 ($-$0.5). 
CPT invariance has been assumed in calculating these elements.
The errors are both statistical and systematic.}
\label{fig:measelems}
\end{figure}

The events are again
divided into eight equal bins of \cthw. Equation~(\ref{eq:proj}) is used to
calculate all six real and three imaginary SDM elements directly.
The  SDM distributions are shown in figure~\ref{fig:measelems}.

This method is used to calculate both the CP-conserving and the CP-violating
 couplings. However, when performing the fit
for CP-conserving couplings only the six real SDM 
coefficients are used, and in the case of CP violating couplings all 
nine SDM coefficients are included. 
The effect of correlations between SDM elements is included.
Although there is no  correlation  between \cthw\ 
bins,
 within each bin the measurements of the SDM elements are highly correlated.
The correlations are obtained directly from the data. 

The Monte Carlo sample used in the fitting procedure is generated for
Standard Model couplings using EXCALIBUR.
All \TGC-dependent four-fermion diagrams are included in this sample. 
SDM distributions for arbitrary couplings are obtained by re-weighting
the fully simulated EXCALIBUR Monte Carlo events. Technically this was
achieved by re-weighting events with the matrix elements from~\cite{operate}.
The five characteristic W production and decay angles are constructed 
from the original four-momenta of the primary fermions in the
simulated EXCALIBUR events. This treatment neglects the effects of
four-fermion background, but these are checked explicitly for
CP-conserving couplings, and found to be negligible.
The expected
contribution from four-fermion background events produced in channels
unaffected by triple gauge couplings, and those from other backgrounds,
are added to the EXCALIBUR sample.

As the SDM elements are normalised to the number of events in each 
$\cos\theta_{\rm W}$ bin, information 
from the production angle of the \W\ is effectively 
removed from the SDM fit. In order to include this 
an independent $\chi^2$ is  derived from the comparison of the shape of 
the $\cos\theta_{\rm W}$ distributions 
and this is incorporated into the fit. No information from the
overall cross-section is included.

The method of re-weighting and fitting is tested by performing fits to 
large samples of fully simulated four-fermion Monte Carlo samples
generated 
with various anomalous couplings as well as others generated with Standard Model
couplings. These bias tests show that the extracted coupling values are 
consistent in all cases with the generated values.
The reliability of the statistical error is also tested by 
performing a fit to a large number of subsamples of the Monte Carlo, each with the
same statistics as the data. It is found that in the case of the Standard Model
 Monte 
Carlo at least 68\% of the fitted values of the couplings fall 
within $\Delta\chi^{2}$=1 of the Standard Model 
values of the couplings. Similar tests are performed on samples of Monte Carlo
with anomalous couplings. Statistics are lower, but in all cases the results
are consistent with the expectation, as seen for the Standard Model samples.

For the CP-violating couplings, including statistical uncertainties and the 
systematic uncertainties described in section~\ref{sec:systematics}, 
we find:
\begin{eqnarray}
\tilde{\kappa}_{\rm z} &=& -0.20^{+0.10}_{-0.07} \nonumber\\ 
g^{\rm z}_{4} &=& -0.02^{+0.32}_{-0.33} \nonumber\\
\tilde{\lambda}_{\rm z} &=& -0.18^{+0.24}_{-0.16} \nonumber
\end{eqnarray}
and for the CP-conserving couplings we find:
\begin{eqnarray}
 \Delta\kappa_{\gamma} &=& -0.22^{+0.29}_{-0.24} \nonumber\\
 \Delta g^{\rm z}_{1} &=& -0.03^{+0.09}_{-0.09} \nonumber\\
 \lambda &=& -0.08^{+0.10}_{-0.09} \nonumber
\end{eqnarray}
These latter results are consistent with those measured in~\cite{189paper}.
All  couplings are set to their Standard Model values except the 
coupling being measured and those related to it via the $SU(2)_{\rm L}\times U(1)_{\rm Y}$ symmetry.
A full breakdown of the results for the CP-violating couplings is shown in 
table~\ref{table:cpvioresults}. 

\begin{figure}[H]
\begin{center}
\epsfig{file=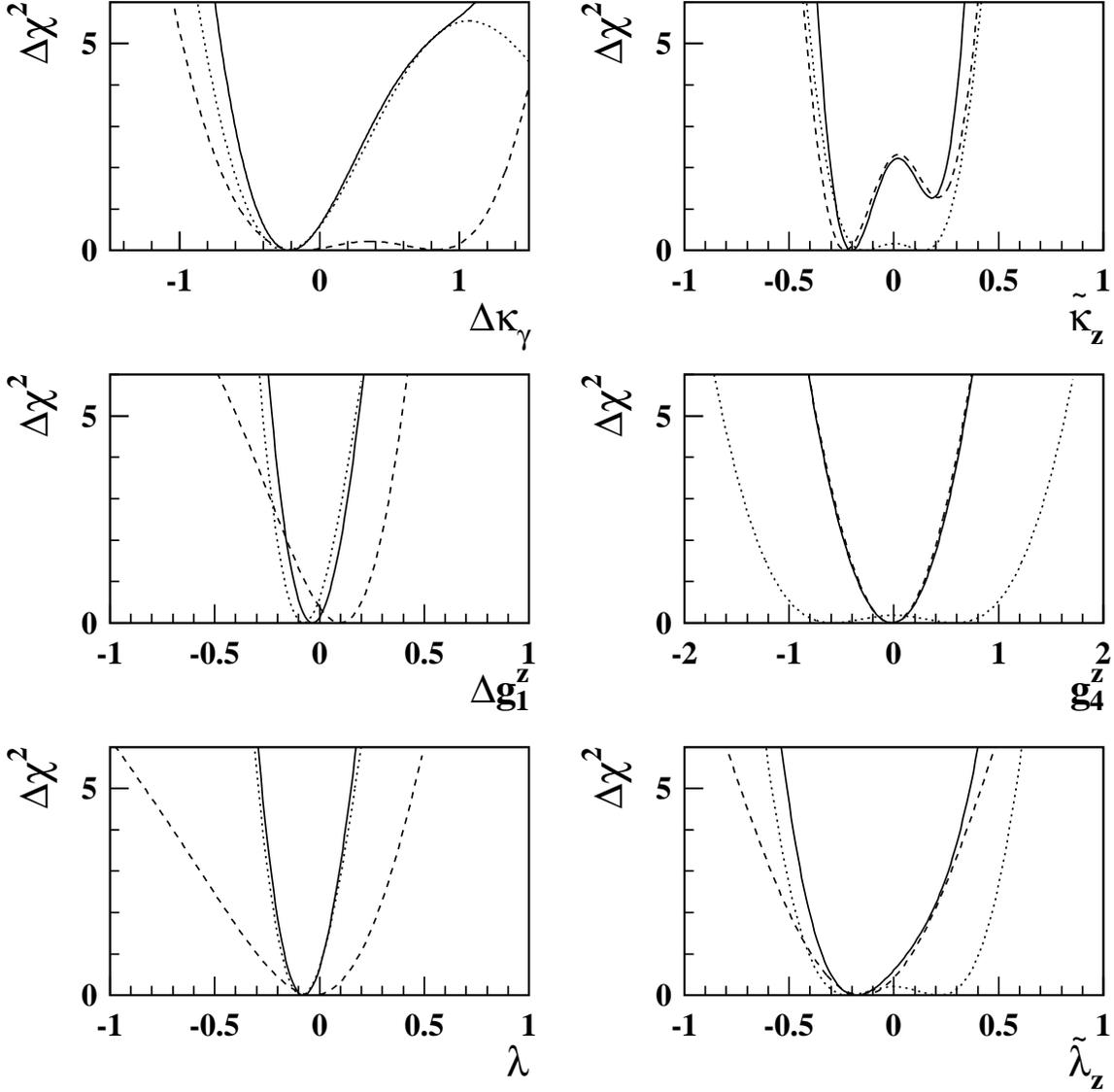,width=1.0\linewidth}
\end{center}
\caption[stuff]{The $\chi^2$ plots for the fits to the 
CP-conserving and CP-violating anomalous couplings. For the CP-conserving couplings the dashed line is the fit to just the 6 real SDM elements, for the CP-violating couplings it is the fit to all 9 SDM elements. The dotted line is the fit to just the 
$\cos\theta_{\rm W}$ distribution. The solid line is the combined fit. All fits include systematic uncertainties. }
\label{fig:chinocp}
\end{figure}

The $\chi^2$ plots for all the fits including
systematics can be seen in figure~{\ref{fig:chinocp}}.
The double minima in the $\chi^2$ curves for
the CP-violating couplings derived from the $\cos\theta_{\rm W}$ distribution
alone reflect that the \cthw\ distribution is sensitive only to the
absolute magnitude of the coupling. 
 The same is
true for the real SDM elements resulting in the double minima
in the 
SDM $\chi^2$ curve for $\tilde{\kappa}_{Z}$. It is only the
imaginary parts that
lift the degeneracy. It is evident that the CP-violating couplings are
better constrained by the SDM elements than by the W boson production
angular distribution, whereas for the CP-conserving couplings the converse
is true.

\begin{figure}[H]
\begin{center}
\epsfig{file=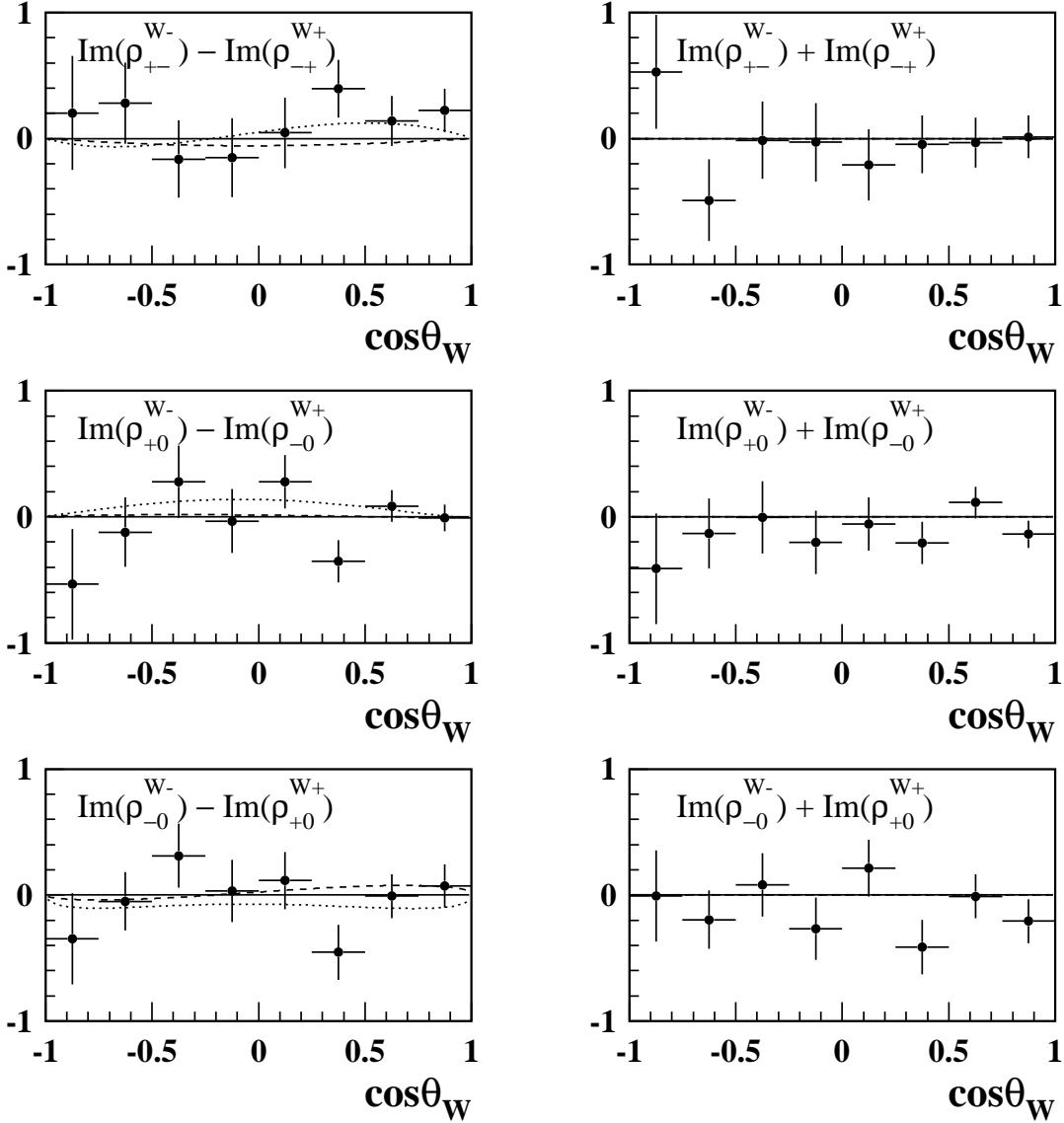,width=1.0\linewidth}
\end{center}
\caption[stuff]{The combination of imaginary SDM coefficients
sensitive to CP-violation at the triple gauge boson vertex for the
189 GeV data (three left side plots).  Overlaid are the Born level 
predictions for the 
Standard Model (solid line) and also anomalous CP-violating
couplings $\tilde{\lambda}_{\rm z} = -0.5$ (dotted line) and $\tilde{\kappa}_{\rm z} = +0.5$ (dashed line). Also shown are the combination of imaginary SDM coefficients
sensitive only to loop effects (any deviation from zero) from the 
189 GeV data (three right side plots). In all cases the errors include both statistical and systematic uncertainties.}
\label{fig:corelmcp}
\end{figure}

A further test of tree level
 CP violation is given by comparing the imaginary SDM elements from the W$^{+}$ 
to those from the W$^{-}$, as in equation~(\ref{eq:test}). These comparisons, 
which
give a model independent test of CP-violation in the triple gauge coupling,
are shown in figure~\ref{fig:corelmcp}. Deviations from zero would only
be due to CP-violation at tree level. No deviations are seen, complementing 
the results of the measurements of the CP-violating 
TGCs.

Also shown in figure~\ref{fig:corelmcp} are the combinations of imaginary SDM
elements that test for effects beyond tree level, as described by 
equation~(\ref{eq:stuff2}).
Any deviation from zero in these plots could only be caused by either loop 
effects or CPT-violation;
all CP-violating tree level effects cancel out. The data are consistent with no 
effect.

\section{Systematic Uncertainties}
\label{sec:systematics}
Systematic uncertainties in the measurements are calculated as described below. The
individual contributions to the errors on the polarised cross-section fractions
 are given in
table~\ref{table:syscross}. For the measurements of the TGCs, all error sources listed below are
included except that from the detector correction. For all TGC measurements
the systematic errors are included as extra uncertainties on the
contents of each bin of the SDM element and \cthw\ distributions. Each
systematic uncertainty is taken to be uncorrelated between bins in both 
\cthw\ and the different SDM element distributions.

$\bullet$ Jet reconstruction: Uncertainties in the modelling of 
jet reconstruction are estimated by varying the reconstruction in a 
Monte Carlo sample. The resolutions of the three jet parameters 
(energy, $\cos\theta_{j}$, $\phi_{j}$) are varied by 10\%, and the 
energy 
is shifted by 5\% to account for systematic uncertainties. 
This is done for both the quark jets and the $\tau$ jets. 
The size of the variations for the quark jets is determined from extensive 
studies of 
back-to-back jets at ${\rm Z}^{0}$ energies. A possible systematic 
shift in the reconstructed direction of the boson has been 
estimated using radiative ${\rm Z}^{0}/\gamma \rightarrow {\rm q}\bar{\rm q}$ events. 
The possible shift in $|\cos\theta_{\rm W}|$ was found to be less than 0.01~\cite{189paper}. The same uncertainties are taken on the $\tau$ jets as for the
quark jets.

$\bullet$ Hadronisation: Uncertainties due to the hadronisation model 
are estimated by comparing Monte Carlo fragmented with HERWIG5.9 to
 Monte Carlo using the JETSET7.4~\cite{pythia} hadronisation scheme. Both 
Monte Carlos have been tuned to OPAL data. Variations in the calculated polarised cross-sections, SDM elements and $\cos\theta_{\rm W}$ distribution between the different samples are taken as the systematic uncertainty.

$\bullet$ Monte Carlo generators: The
 modelling of the four-fermion production processes is assessed by
comparing SDM elements and \cthw\ distributions from the ERATO and grc4f 
Monte Carlo programs to those from  EXCALIBUR. All these generators have a different calculation of the matrix elements and a different treatment of ISR. 
The calculated values of the individual W and W-pair polarised cross-section 
fractions are also compared. Variations in the calculated polarised cross-sections, SDM elements and $\cos\theta_{\rm W}$ distribution between the different samples are taken as the systematic uncertainty.

$\bullet$ Background simulation: Possible systematic effects 
due to the simulation of the dominant ${\rm Z}^{0}/\gamma$ 
background are accounted for by replacing the PYTHIA Monte 
Carlo with a HERWIG sample. The two-photon background is removed and doubled
as well. 
Variations in the extracted polarised cross-sections, SDM elements and $\cos\theta_{\rm W}$ distribution are taken as the systematic uncertainty.

$\bullet$ Lepton ID: The efficiency of the lepton identification in the data compared to that in the Monte Carlo is investigated as a function of the polar angle and energy of the lepton. The Monte Carlo is weighted to adjust these distributions to the data. Variations in the calculated polarised cross-sections, SDM elements and $\cos\theta_{\rm W}$ distribution between the  samples before and after weighting are taken as the systematic uncertainty.

$\bullet$ Detector effect correction: The use of Standard Model Monte Carlo to
correct for detector effects could introduce some bias. This is tested for by comparing the helicity fractions 
         calculated from generator level non-Standard Model Monte Carlo
         with those calculated from the same sample after full detector
         simulation and correction using correction factors determined
         from Standard Model Monte Carlo. 
This test is done with six samples of Monte Carlo each
with one of the couplings $\Delta\kappa_{\gamma}$, $\Delta g^{z}_{1}$, $\lambda$, set to $\pm1$.

$\bullet$ Lepton energy scale and charge misassignment: Uncertainties on the lepton energy scale are tested by shifting the lepton energy by 0.3\%. These are found to be negligible. The momentum resolution of electrons and muons and their charge
misassignment are investigated by varying the resolution in Q/p$_{\rm t}$ 
by 10\%. Here Q is the lepton charge and p$_{\rm t}$  is the transverse momentum
with respect to the beam direction.

\section{Conclusion}

W-pair events with one leptonically and one hadronically decaying W are
analysed to extract the polarisation properties of the W boson and test
CP invariance. Both the individual W, and W-pair, polarised cross-sections
have been measured. The individual W polarised cross-sections are
well-described by the Standard Model predictions, both inclusively and
separately for the leptonically and hadronically decaying W bosons 
(table~\ref{table:hel}). The results are consistent with those measured by  L3
 at the same centre-of-mass energy~\cite{l3}, with the results presented here
being more precise.

The W-pair polarised cross-sections are measured for the first time at 
LEP. All the results are found to be consistent with the
Standard Model expectations (table~\ref{table:hel2}).

Triple gauge couplings are extracted using information from the
leptonically decaying W, together with the W production polar angle.
They are found to be consistent with the Standard Model predictions. 
The CP conserving couplings have been measured elsewhere with the same
data sample~\cite{189paper}, and serve as a consistency check for this 
analysis. 
The CP violating couplings $\tilde{\kappa}_{\rm z}$, $\tilde{\lambda}_{\rm z}$ and 
$g^{\rm z}_{4}$ are measured for the first
time with OPAL data, table~\ref{table:cpvioresults}, and constraints are thus 
placed on possible new
CP-violating contributions to the W production and decay processes.

\section*{Acknowledgements}

We particularly wish to thank the SL Division for the efficient operation
of the LEP accelerator at all energies
 and for their continuing close cooperation with
our experimental group.  We thank our colleagues from CEA, DAPNIA/SPP,
CE-Saclay for their efforts over the years on the time-of-flight and trigger
systems which we continue to use.  In addition to the support staff at our own
institutions we are pleased to acknowledge the  \\
Department of Energy, USA, \\
National Science Foundation, USA, \\
Particle Physics and Astronomy Research Council, UK, \\
Natural Sciences and Engineering Research Council, Canada, \\
Israel Science Foundation, administered by the Israel
Academy of Science and Humanities, \\
Minerva Gesellschaft, \\
Benoziyo Center for High Energy Physics,\\
Japanese Ministry of Education, Science and Culture (the
Monbusho) and a grant under the Monbusho International
Science Research Program,\\
Japanese Society for the Promotion of Science (JSPS),\\
German Israeli Bi-national Science Foundation (GIF), \\
Bundesministerium f\"ur Bildung und Forschung, Germany, \\
National Research Council of Canada, \\
Research Corporation, USA,\\
Hungarian Foundation for Scientific Research, OTKA T-029328, 
T023793 and OTKA F-023259.\\

\begin{center}
\begin{table}[H]
\begin{center}
\begin{tabular}{|l|c|c|c|} \hline
Fit & $\tilde{\kappa}_{\rm z}$ & $g^{\rm z}_{4}$ & $\tilde{\lambda}_{\rm z}$ \\\hline
 & & & \\
SDM Elements & $-0.19^{+0.08}_{-0.07}$  & $\;\;\;$$0.00^{+0.21}_{-0.20}$ & $-0.12^{+0.17}_{-0.16}$  \\
 & & & \\
$\cos\theta_{\rm W}$ & $-0.19^{+0.46}_{-0.08}$ & $\;\;\;$$0.7^{+0.4}_{-1.8}$ & $-0.29^{+0.69}_{-0.11}$  \\\hline
 & & & \\
Combined & $-0.19^{+0.06}_{-0.05}$ & $\;\;\;$$0.01^{+0.22}_{-0.22}$ & $-0.19^{+0.18}_{-0.13}$  \\
 & & &  \\
Expected Stat. Error  & $\;\;\;\pm$0.11 & $\;\;\;\pm$0.19 & $\;\;\;\pm$0.12   \\\hline
 & & &  \\
Final Fit    & $-0.20^{+0.10}_{-0.07}$ & $-0.02^{+0.32}_{-0.33}$ & $-0.18^{+0.24}_{-0.16}$ \\
Including Systematics & & & \\\hline
\end{tabular}
\caption{Measured values of the CP-violating TGC parameters. Both the SDM elements for the leptonically decaying W and the $\cos\theta_{\rm W}$ production distribution in ${\rm q}{\rm q}\ell{\nu}$ events from the 189 GeV data are used in the calculation. Errors are statistical only except in the case of the final combined fit.}
\label{table:cpvioresults}
\end{center}
\end{table}
\end{center}

\begin{center}
\begin{table}[H]
\begin{center}
\begin{tabular}{|l|c|c|c|c|} \hline
Systematic & TT & LL & TL & T,L \\\hline 
\underline{Jet Reconstruction} & & & &\\
WW$\rightarrow{\rm q}{\rm q}\ell{\nu}$  & 0.013 & 0.004 & 0.008 & 0.007 \\
W$\rightarrow\ell{\nu}$ & - & - & - & 0.005 \\
W$\rightarrow{\rm q}{\rm q}$ & - & - & - & 0.011 \\\hline
\underline{Hadronisatation} & & & &\\
WW$\rightarrow{\rm q}{\rm q}\ell{\nu}$  & 0.016 & 0.002 & 0.014 & 0.009 \\
W$\rightarrow\ell{\nu}$ & - & - & - & 0.003\\
W$\rightarrow{\rm q}{\rm q}$ & - & - & - & 0.021\\\hline
\underline{MC Generator} & & & &\\
WW$\rightarrow{\rm q}{\rm q}\ell{\nu}$  & 0.015 & 0.013 & 0.028 & 0.001\\
W$\rightarrow\ell{\nu}$ & - & - & - & 0.004 \\
W$\rightarrow{\rm q}{\rm q}$ & - & - & - & 0.006 \\\hline
\underline{Background} & & & & \\
WW$\rightarrow{\rm q}{\rm q}\ell{\nu}$  & 0.004 & 0.001 & 0.004 & 0.002 \\
W$\rightarrow\ell{\nu}$ & - & - & - & 0.003 \\
W$\rightarrow{\rm q}{\rm q}$ & - & - & - & 0.004\\\hline
\underline{Lepton id} & & & & \\
WW$\rightarrow{\rm q}{\rm q}\ell{\nu}$  & 0.017 & 0.005 & 0.014 & 0.010\\
W$\rightarrow\ell{\nu}$ & - & - & - & 0.017 \\
W$\rightarrow{\rm q}{\rm q}$ & - & - & - & 0.003\\\hline
\underline{Detector Effect Correction} & & & &  \\
WW$\rightarrow{\rm q}{\rm q}\ell{\nu}$  & 0.004 & 0.008 & 0.012 & 0.002 \\
W$\rightarrow\ell{\nu}$ & - & - & - & 0.009 \\
W$\rightarrow{\rm q}{\rm q}$ & - & - & - & 0.005\\\hline
\underline{Charge Misassignment} & & & &  \\
WW$\rightarrow{\rm q}{\rm q}\ell{\nu}$  & 0.007 & 0.001 & 0.008 & 0.005 \\
W$\rightarrow\ell{\nu}$ & - & - & - & 0.011 \\
W$\rightarrow{\rm q}{\rm q}$ & - & - & - & 0.001\\\hline
\underline{Total} & & & &  \\
WW$\rightarrow{\rm q}{\rm q}\ell{\nu}$  & 0.032 & 0.017 & 0.038 & 0.016\\
W$\rightarrow\ell{\nu}$ & - & - & - & 0.023 \\
W$\rightarrow{\rm q}{\rm q}$ & - & - & - & 0.025\\\hline
\end{tabular}
\caption{The contribution to the systematic uncertainty on the polarised cross-section fractions from different sources.}
\label{table:syscross}
\end{center}
\end{table}
\end{center}

\end{document}